\documentclass[twocolumn]{revtex4}
\usepackage{}
\usepackage{epsfig,latexsym,amssymb}
\usepackage{latexsym}
\usepackage{amsmath}
\usepackage{amssymb}
\usepackage{amsfonts}
\usepackage{comment}
\usepackage{graphicx}
\usepackage{verbatim}
\usepackage{epstopdf}
\usepackage{hyperref, url}
\hypersetup{colorlinks   = true,
            urlcolor     = blue,
            citecolor    = blue,
            linkcolor    = blue,
            menucolor    = blue,
            anchorcolor  = blue,
            filecolor    = blue}

\newcommand{\be}{\begin{equation}}
\newcommand{\ee}{\end{equation}}
\newcommand{\bea}{\begin{eqnarray}}
\newcommand{\eea}{\end{eqnarray}}

\begin{document}

\title{Stimulated decay of collapsing axion stars and fast radio bursts}

\author{Haoran Di \footnote{hrdi@ecut.edu.cn}}
\affiliation{School of Science, East China University of Technology, Nanchang 330013, China}

\begin{abstract}
The radiation mechanism of fast radio bursts (FRBs) has been extensively studied but still remains elusive. In the search for dark matter candidates, the QCD axion and axionlike particles (ALPs) have emerged as prominent possibilities. These elusive particles can aggregate into dense structures called axion stars through Bose-Einstein condensation (BEC). Such axion stars could constitute a significant portion of the mysterious dark matter in the universe. When these axion stars grow beyond a critical mass, usually through processes like accretion or merging, they undergo a self-driven collapse. Traditionally, for spherically symmetric axion clumps, the interaction between axions and photons does not lead to parametric resonance, especially when the QCD axion-photon coupling is at standard levels.
Nevertheless, our study indicates that even QCD axion stars with typical coupling values can trigger stimulated decay during their collapse, rather than producing relativistic axions through self-interactions. This process results in short radio bursts, with durations of around 0.1 seconds, and can be potentially observed using radio telescopes like FAST or SKA. Furthermore, we find that collapsing axion stars for ALPs with specific parameters may emit radio bursts lasting just milliseconds with a peak luminosity of $1.60\times10^{42}\rm{erg/s}$, matching the characteristics of the observed non-repeating FRBs.
\end{abstract}

\maketitle

\section{Introduction}
The accumulation of evidence from diverse observations and theoretical models has resulted in the understanding that dark matter makes up a significant fraction of the overall energy density. Nevertheless, the nature and composition of dark matter particles continue to be enigmatic and unknown. The QCD axion \cite{Weinberg:1977ma,Wilczek:1977pj}, a candidate for dark matter, is a product of the Peccei-Quinn mechanism \cite{Peccei:1977ur,Peccei:1977hh}, which serves as a prominent solution to the strong-CP problem.
Furthermore, within the framework of string theory, there are strong incentives for the existence of a diverse array of axionlike particles (ALPs) covering a broad spectrum of mass scales, leading to the concept known as the ``axiverse" \cite{Arvanitaki:2009fg}. In this paper, we will use the term ``axions" to refer to both QCD axions and ALPs. Axions can arise through various mechanisms, such as the misalignment mechanism \cite{Preskill:1982cy,Abbott:1982af,Dine:1982ah}, string defects \cite{Gorghetto:2020qws}, or the kinetic misalignment mechanism \cite{Co:2019jts}, among others. Due to their bosonic characteristics, axions have the capacity to achieve remarkably high phase space density, resulting in the fascinating phenomenon of Bose-Einstein condensation (BEC) \cite{Sikivie:2009qn}. Due to BEC, axions have the capability to come together and form gravitationally bound entities recognized as axion stars \cite{Braaten:2019knj,Visinelli:2021uve}.

As the axion condensate undergoes coherent oscillations, it has the potential to induce parametric resonance \cite{Hertzberg:2018zte,Arza:2018dcy,Arza:2020eik,Levkov:2020txo,Chung-Jukko:2023cow} in photons, resulting in an exponential increase in photon population and subsequent radio wave emission. In the case of spherically symmetric axion stars, resonance is typically not observed for the usual values of the QCD axion-photon coupling unless it involves relatively higher couplings or QCD axion stars with substantial angular momentum \cite{Hertzberg:2018zte}.
Additionally, when the axion-photon coupling surpasses the original Kim-Shifman-Vainshtein-Zakharov (KSVZ) value by two orders of magnitude, collapsing QCD axion stars produce radio bursts \cite{Levkov:2020txo}.

Fast radio bursts (FRBs) are bright transient radio signals that endure for just milliseconds \cite{Lorimer:2007,Keane:2012yh,Thornton:2013iua}, and their physical origin, as well as the mechanism of radiation, remain shrouded in mystery. Thus far, hundreds of FRBs have been detected, with a limited subset demonstrating repeating behaviors \cite{Spitler:2016dmz,CHIMEFRB:2019pgo,Fonseca:2020cdd,CHIMEFRB:2023myn}. It is commonly believed that repeating FRBs and non-repeating FRBs arise from different physical processes. Numerous models have been put forward to elucidate the sources of FRBs. See Refs. \cite{Katz:2018xiu,Popov:2018hkz,Platts:2018hiy,Petroff:2019tty,Cordes:2019cmq,Zhang:2020qgp,Xiao:2021omr} for the review of FRBs. The frequency spectrum of FRBs typically falls within the range of approximately $400\rm{MHz}$ to $8\rm{GHz}$ \cite{Petroff:2019tty}, and the total energy emitted typically ranges from $10^{38}$ to $10^{40}\rm{erg}$ \cite{Thornton:2013iua,Spitler:2014fla}. While a solid link between as least some FRBs and magnetars \cite{CHIMEFRB:2020abu,Bochenek:2020zxn} has been confirmed, the mechanisms triggering these enigmatic phenomena and their radiation processes remain subjects of intense controversy.

In this paper we propose that stimulated decay \cite{Kephart:1994uy,Rosa:2017ury,Caputo:2018vmy} of collapsing dilute axion stars for ALPs with smaller values of the axion-photon coupling compared with QCD axion may be the origin of some of the observed non-repeating FRBs. See Refs. \cite{Iwazaki:2014wka,Tkachev:2014dpa,Raby:2016deh,Buckley:2020fmh} for relevant works about axion stars and FRBs.
In this paper we use the natural units, $c = \hbar=1$.

\section{Dilute Axion Stars}

The QCD axion \cite{Weinberg:1977ma}\cite{Wilczek:1977pj} is a pseudoscalar boson with a spin of 0, characterized by its small mass referred to as  $m_\phi$, extraordinarily weak self-interaction, and highly feeble interactions with particles in the Standard Model.
The demand for Lagrangian shift symmetry invariance implies that the axion potential $V(\phi)$ must display periodic characteristics in relation to $\phi$:
\bea
V(\phi)=V(\phi+2\pi f_a),
\eea
where $f_a$ is denoted as the axion decay constant, which signifies the energy scale at which the spontaneous breaking of the $U(1)$ symmetry occurs. The most commonly employed model for the axion potential in the majority of phenomenological investigations is the instanton potential \cite{Peccei:1977ur}:
\bea
V(\phi)&=&(m_\phi f_a)^2\left[1-\cos\left(\frac{\phi}{f_a}\right)\right]\\ \nonumber
&=&{1\over2} m_\phi^2 \phi^2+\frac{\lambda}{4!}\phi^4+...,
\eea
where $\lambda=-m_\phi^2/f_a^2$ is the attractive coupling of self-interaction.
Axions, being bosons, are capable of achieving exceedingly high phase space density, leading to the formation of BECs \cite{Sikivie:2009qn}. These BECs can lead to the formation of axion stars, which may vary in density, existing in both dilute and dense configurations \cite{Chavanis:2017loo,Visinelli:2017ooc,Eby:2019ntd}. However, the duration of a dense axion star's existence could be too short to confer it cosmological importance as an astrophysical entity \cite{Braaten:2019knj,Seidel:1991zh,Hertzberg:2010yz,Eby:2015hyx}.

A stable dilute axion star configuration can be conceptualized as an equilibrium between the attractive self-gravity of axion particles and the repulsive gradient energy. This equilibrium can persist as long as the star maintains a sufficiently low density, making the influence of self-interactions insignificant.
The maximum mass \cite{Chavanis:2011zi}\cite{Chavanis:2011zm} and the corresponding radius of a dilute axion star are given by
\bea
\label{maximum mass}
M_{\rm{max}}\sim 5.073\frac{M_{pl}}{\sqrt {|\lambda|}},~~~~~~R_{\rm{min}}\sim\sqrt {|\lambda|}\frac{M_{pl}}{m_\phi}\lambda_c,
\eea
where $M_{pl}$ is the Planck mass and $\lambda_c$ is the Compton wavelength of axion.
When the axion star's mass increases and exceeds the critical mass given by Eq.\eqref{maximum mass} due to merger events \cite{Mundim:2010hi,Cotner:2016aaq,Schwabe:2016rze,Eby:2017xaw,Hertzberg:2020dbk,Du:2023jxh} or the accretion of axions from the background \cite{Chen:2020cef,Chan:2022bkz,Dmitriev:2023ipv}, self-interactions become significant and may lead to the destabilization of the star.

As the axion star initiates its collapse, there is a rapid escalation in its density. When the star's size approaches the Compton wavelength of the axion, $2\pi/m_\phi$, the axions undergo annihilations, transitioning into relativistic states \cite{Levkov:2016rkk}. This process leads to a swift depletion of the collapsing star's energy, a phenomenon commonly referred to as ``bosenova'' \cite{Chavanis:2016dab,Levkov:2016rkk,Eby:2016cnq,Fox:2023aat}.

In the traditional post-inflationary model, where the U(1) symmetry spontaneously breaks after the inflationary epoch, QCD axion stars have the potential to constitute up to $75\%$ of the dark matter component \cite{Eggemeier:2019khm,Xiao:2021nkb,Eggemeier:2022hqa}. Furthermore, the analysis of microlensing events in HSC and OGLE data consistently suggests that approximately $27^{+7}_{-13}$ percent of dark matter could be present as axion stars \cite{Sugiyama:2021xqg}.

\section{Spontaneous Decay from Dilute Axion Stars}

In the context of the instanton potential, we can express the general Lagrangian of the axion as follows:
\bea \label{axion}
{\cal L}&=&{1\over 2}\partial_{\mu}\phi\partial^{\mu}\phi-{m_{\phi}^2\over 2}\phi^2
+\frac{\lambda}{4!}\phi^4 \\ \nonumber
&&+{1 \over4}g_{a\gamma\gamma}\phi F_{\mu\nu} \tilde F^{\mu\nu}+...,
\eea
where $F_{\mu\nu}$ represents the electromagnetic field tensor, $\tilde F^{\mu\nu}$ is the dual tensor of $F_{\mu\nu}$ defined as $\frac{1}{2} F_{\alpha\beta} \epsilon^{\mu\nu\alpha\beta}$, $\phi$ is the axions field, and the axion-photon coupling $g_{a\gamma\gamma}$ is given by $g_{a\gamma\gamma}=\alpha K/(2\pi f_a)$, where $\alpha$ is the fine structure constant, and $K$ is a model-dependent constant of order one. For instance, in the standard KSVZ model \cite{Kim:1979if}\cite{Shifman:1979if}, $K$ is approximately $-1.95$, while in the DFSZ model \cite{Dine:1981rt}\cite{Zhitnitsky:1980tq}, $K$ is roughly $0.72$. Therefore, we will set $K=1$ in the following discussion. For QCD axions, there is a relationship between the decay constant $f_a$ and the axion's mass \cite{Sikivie:2006ni}:
\bea \label{relation}
f_a\simeq6\times10^{12}\left(\frac{10^{-6}{\rm{eV}}}{m_{\phi}}\right)\rm{GeV}.
\eea
The current relic density of axions \cite{Di:2023xaw} is described by the equation:
\bea \label{abundance}
\Omega_\phi h^2&\simeq&0.12
\left(\frac{g_\star(T_{\rm{osc}})}{106.75}\right)^{3/4}
\left(\frac{m_\phi}{10^{-6}\rm{eV}}\right)^{1/2}\nonumber\\
&&\times\left(\frac{f_a}{5.32\times 10^{12}\rm{GeV}}\right)^2,
\eea
where $\Omega_\phi=\rho_\phi/\rho_{cr}$ represents the ratio of the energy density of the axion to the critical density. This equation, applicable to axions generated via the misalignment mechanism, sets the lower limit for the axion parameter space, as depicted in Fig.~\ref{fig:constraints}.

Substituting the attractive coupling of self-interaction $\lambda=-m_\phi^2/f_a^2$ into Eq. \eqref{maximum mass} yields
\bea\label{critical mass}
M_{\rm{max}}\sim5.97\times10^{-11}M_{\odot}\left(\frac{m_\phi}{10^{-6}{\rm{eV}}}\right)^{-1}
\left(\frac{f_a}{10^{12}{\rm{GeV}}}\right),
\eea
\bea\label{radius}
R_{\rm{min}}\sim2.41\times10^3{\rm{km}}\left(\frac{m_\phi}{10^{-6}{\rm{eV}}}\right)^{-1}
\left(\frac{f_a}{10^{12}{\rm{GeV}}}\right)^{-1}.
\eea
For general axion stars, we use $M_{\rm{AS}}$ and $R_{\rm{AS}}$ to represent the mass and radius respectively.
By combining Eq. \eqref{relation} with Eq. \eqref{abundance}, we can derive the axion's mass, $m_\phi$, which is approximately $1.17\times10^{-6}{\rm{eV}}$, and the decay constant, $f_a$, which is approximately $5.11\times10^{12}{\rm{GeV}}$, assuming that the primary component of dark matter consists of QCD axions. Substituting these values into Eq. \eqref{critical mass}, we can determine the maximum mass of the axion star to be approximately $5.24\times10^{-10}M_\odot$.

Axions are not entirely stable, primarily due to the interaction term ${\cal L}_{int}=1/4 g_{a\gamma\gamma}\phi F_{\mu\nu} \tilde F^{\mu\nu}$. This interaction leads to the decay of axions into two photons, and the associated decay rate is given by
\bea
\Gamma_\phi&=&\frac{\alpha^2 m_\phi^3}{256 \pi^3 f_a^2}\nonumber\\
&=&1.02\times 10^{-53} {\rm{s}^{-1}} \left( \frac{m_\phi}{10^{-6}\rm{eV}} \right)^3 \left(\frac{10^{12}\rm{GeV}}{f_a}\right)^2.
\eea
This implies that the axion has a finite lifetime determined by the decay rate and the lifetime of axion is characterized by the following expression:
\bea
\tau_\phi=9.80\times 10^{52} {\rm{s}} \left( \frac{10^{-6}\rm{eV}} {m_\phi} \right)^3 \left(\frac{f_a} {10^{12}\rm{GeV}}\right)^2.
\eea
Considering the relationship $f_a\simeq6\times10^{12}(10^{-6}{\rm{eV}}/m_{\phi})\rm{GeV}$, the lifetime of QCD axions becomes
$\tau_\phi=3.53\times 10^{54} {\rm{s}} \left( {m_\phi}/{10^{-6}\rm{eV}}  \right)^{-5}$.
For axions to be viable candidates for dark matter, their lifetimes must exceed the age of the universe. This imposes a constraint on the mass of axions, with $m_\phi\lesssim \rm{few~eV}$ being a requirement. Therefore, for QCD axions with a mass of approximately $1.17\times10^{-6}{\rm{eV}}$, the spontaneous decay of axions would not destabilize dilute axion stars.

\section{Stimulated Decay from Collapsing Axion Stars}

As the axion is a boson, we must take into account the boson enhancement effect. The change in photon number density within the axion star resulting from axion decays and inverse decays is described by the Boltzmann equation \cite{Kephart:1994uy}
\bea
\frac{dn_\lambda}{dt}&=&\int\frac{d^3p}{(2\pi)^32p^0} \int\frac{d^3k_1}{(2\pi)^32k_1^0} \int\frac{d^3k_2}{(2\pi)^32k_2^0}\nonumber\\
 &&\times(2\pi)^4\delta^4(p-k_1-k_2)|{\cal M}|^2\nonumber\\
&&\times\{f_\phi(\textbf{p})[1+f_\lambda(\textbf{k}_1)][1+f_\lambda(\textbf{k}_2)]\nonumber\\
&&-f_\lambda(\textbf{k}_1)f_\lambda(\textbf{k}_2)[1+f_\phi(\textbf{p})]\},
\eea
where $f_i$ denotes the phase space densities of each species, and $n_i$ is the resulting number density, calculated as $n_i=\int d^3 k/(2\pi)^3f_i$. $\cal M$ stands for the matrix element associated with the interaction term ${\cal L}_{int}=1/4 g_{a\gamma\gamma}\phi F_{\mu\nu} \tilde F^{\mu\nu}$. For the purpose of solving this equation, we make the assumption that the phase space distribution of axions and photons is roughly uniform and isotropic within the axion star. This assumption leads to the following expression for the total photon number density \cite{Rosa:2017ury}:
\bea\label{photon number density}
\frac{dn_\gamma}{dt}=\Gamma_\phi[2n_\phi
\left(1+\frac{8\pi^2}{m_\phi^3v}n_\gamma\right)
-\frac{16\pi^2}{3m_\phi^3}
\left(v+\frac{3}{2}\right)n_\gamma^2],
\eea
where $v$ represents the maximum axion velocity within the axion star, which is approximately equal to $1/(2R_{\rm{AS}}m_\phi)$ as determined by the Heisenberg uncertainty principle.
\begin{figure}
\begin{center}
\includegraphics[width=0.45\textwidth]{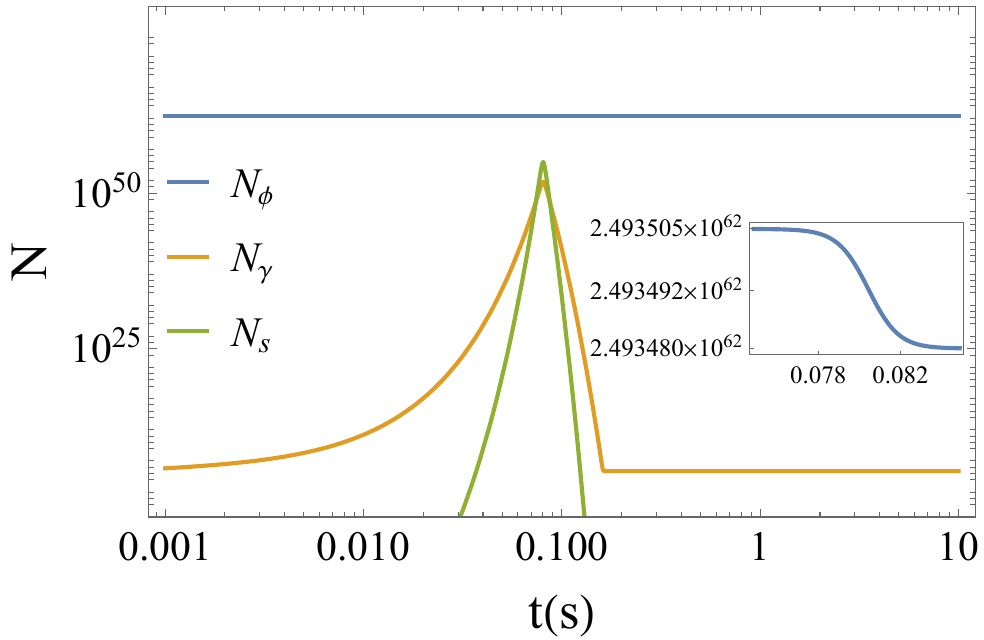}
\caption{Numerical evolution of the axion, photon, and ``sterile'' axion numbers for a collapsing axion star below the critical radius $R_{cr}$, where stimulated radiation begins. The inset plot magnifies the evolution of axion numbers near the peak of stimulated radiation. The axion has a mass of $m_\phi=1.17\times10^{-6}{\rm{eV}}$, and the decay constant is $f_a=5.11\times10^{12}{\rm{GeV}}$. The time when stimulated radiation begins is set as the starting point for timing.}
\label{fig:number}
\end{center}
\end{figure}

\begin{figure}
\begin{center}
\includegraphics[width=0.45\textwidth]{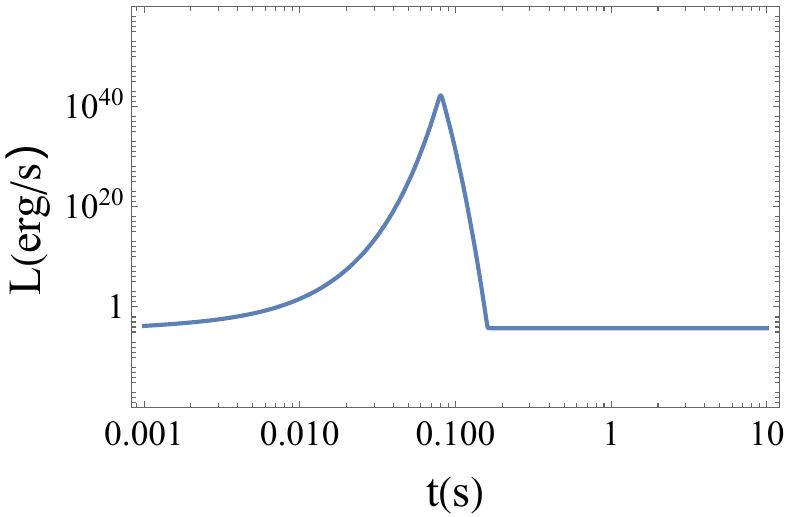}
\caption{Luminosity of stimulated decay in axion stars with $m_\phi=1.17\times10^{-6}{\rm{eV}}$ and decay constant $f_a=5.11\times10^{12}{\rm{GeV}}$. The peak luminosity is approximately $10^{42}\rm{erg/s}$, and the total energy emitted by this radio burst is around $10^{38}\rm{erg}$.}
\label{fig:luminosity}
\end{center}
\end{figure}
In this equation, the initial terms enclosed within the square brackets account for spontaneous and stimulated decay, with the latter being directly proportional to axion number density $n_\phi$ and photon number density $n_\gamma$. The final terms, which are proportional to $n_\gamma^2$, represent the process of inverse decay. The term in proportion to $v$ denotes annihilation, resulting in a low-energy ``active" axion that can remain within the axion star. Conversely, the term scaled by a factor of $3/2$ signifies the production of ``sterile" axions \cite{Kephart:1994uy}, which escape from the axion star.
Furthermore, the photons produced by spontaneous and stimulated decay escape from the axion star at a rate
\bea \label{escape rate}
\Gamma_e=\frac{1}{R_{\rm{AS}}},
\eea
which is the inverse of the axion star's light-crossing time. By substituting Eq. \eqref{radius} into Eq. \eqref{escape rate}, we obtain
\bea
\Gamma_e\sim1.24\times10^2{\rm{s^{-1}}}\left(\frac{m_\phi}{10^{-6}{\rm{eV}}}\right)
\left(\frac{f_a}{10^{12}{\rm{GeV}}}\right)
\eea

\begin{figure}
\begin{center}
\includegraphics[width=0.45\textwidth]{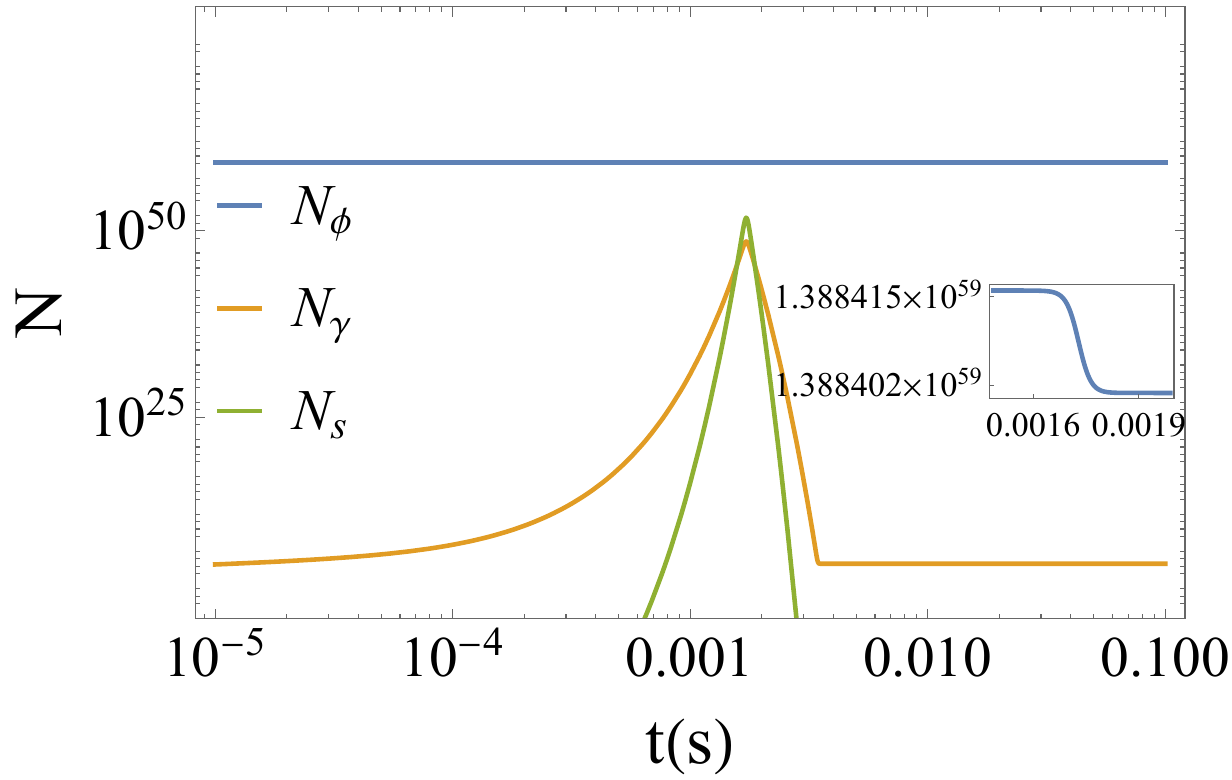}
\caption{Numerical evolution of the axion, photon, and ``sterile'' axion numbers for a collapsing axion star below the critical radius $R_{cr}$, where stimulated radiation begins, with $m_\phi=5\times10^{-5}{\rm{eV}}$ and decay constant $f_a=5.20\times10^{12}{\rm{GeV}}$. The burst has a duration of approximately a few milliseconds and a peak luminosity of $1.60\times10^{42}\rm{erg/s}$.}
\label{fig:number2}
\end{center}
\end{figure}

\begin{figure}
\begin{center}
\includegraphics[width=0.45\textwidth]{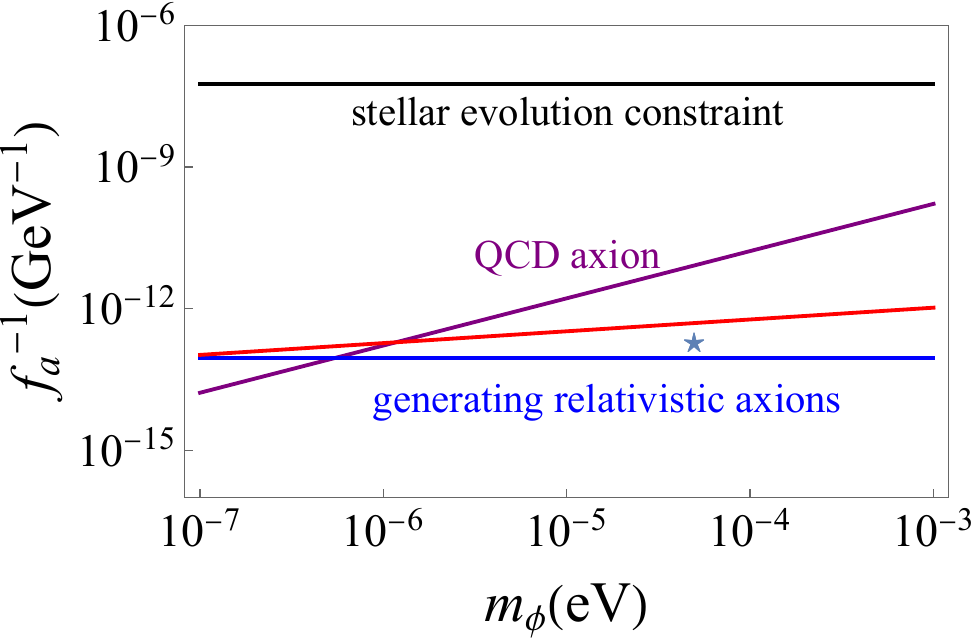}
\caption{Constraints on the parameter space of axions. The parameter space above the black solid line is excluded by the constraints of stellar evolution \cite{Ayala:2014pea,Giannotti:2015kwo}. The purple line represents QCD axions, and the red line represents the cosmological abundance of axions, as given by Eq.\eqref{abundance}, produced by the misalignment mechanism. In the area below the blue line, collapsed axion stars will emit relativistic axions \cite{Levkov:2016rkk} instead of radio bursts. The parameter space around the ``star'' in the figure can explain the non-repeating FRBs. Interestingly, this ``star'' happens to be near the cosmological abundance of axions.}
\label{fig:constraints}
\end{center}
\end{figure}

Therefore, upon integrating Eq. \eqref{photon number density} and considering escaped photons, we derive the following system of coupled differential equations governing the evolution of the number of axions and photons within the axion star:
\bea
\frac{dN_\gamma}{dt}=\Gamma_\phi[2N_\phi(1+AN_\gamma)-BN_\gamma^2]-\Gamma_e N_\gamma,
\eea
\bea
\frac{dN_\phi}{dt}=-\Gamma_\phi[N_\phi(1+AN_\gamma)-CN_\gamma^2],
\eea
where $A=6\pi(m_\phi^3vR_{\rm{AS}}^3)^{-1}$, $B=4\pi (v+3/2)(m_\phi^3R_{\rm{AS}}^3)^{-1}$, $C=4\pi v(m_\phi^3R_{\rm{AS}}^3)^{-1}$.

For maximum mass dilute axion stars, the photon number $N_\gamma\simeq(2\Gamma_\phi/\Gamma_e)N_\phi$ present in axion stars due to spontaneous decay are not sufficient to trigger stimulated decay until the axion star's size is less than the critical radius $R_{cr}\sim24\pi\Gamma_\phi N_\phi/m_\phi^2$, i.e. $N_\gamma A\sim1$, stimulated decay begins. The spectrum line is a nearly monochromatic frequency $f\simeq m_\phi/({4\pi})\approx1.21(m_\phi/10^{-5}{\rm{eV}}){\rm{GHz}}$ which is a obvious characteristic signal. However, when the star's size approaches the Compton wavelength of the axion, $2\pi/m_\phi$, the axions undergo annihilations, transitioning into relativistic states \cite{Levkov:2016rkk}.
Therefore, the critical radius $R_{cr}\sim24\pi\Gamma_\phi N_\phi/m_\phi^2$ should greater than the Compton wavelength $2\pi/m_\phi$ in order to induce stimulated radiation instead of generating relativistic axions, i.e. $f_a<1.08\times10^{13}\rm{GeV}$, as shown in Fig.~\ref{fig:constraints}.

In Fig.~\ref{fig:number},  we present a numerical solution for a representative set of parameters, where $m_\phi=1.17\times10^{-6}{\rm{eV}}$ and the decay constant $f_a=5.11\times10^{12}{\rm{GeV}}$. Once the size of the axion star falls below the critical radius, it will generate a radio burst with a duration of approximately 0.1 seconds, which can be seen from Fig.~\ref{fig:number}.
The luminosity of the collapsing axion star is given by:
\bea
L_\phi=\Gamma_e m_\phi N_\gamma/2,
\eea
which is illustrated in Fig.~\ref{fig:luminosity} for the values $m_\phi=1.17\times10^{-6}{\rm{eV}}$ and the decay constant $f_a=5.11\times10^{12}{\rm{GeV}}$. The peak luminosity reaches approximately $10^{42}\rm{erg/s}$, and the total energy released by this radio burst amounts to around $10^{38}\rm{erg}$. The corresponding photon frequency is $141.57\rm{MHz}$, making it detectable by radio telescopes like the Five-hundred-meter Aperture Spherical radio Telescope (FAST) within its frequency range of $0.1-3\rm{GHz}$ or the Square Kilometer Array (SKA) with a broader frequency range of $0.05-14\rm{GHz}$ \cite{SKA:2015}.
Therefore, a collapsing QCD axion star with an axion mass $m_\phi\gtrsim10^{-6}{\rm{eV}}$ will initiate stimulated decay without requiring the axion-photon coupling to exceed the original KSVZ value by two orders of magnitude. This radio burst, lasting for about 0.1 seconds, is potentially detectable by radio telescopes like FAST or SKA.

The collapsing axion star for ALPs with $m_\phi=5\times10^{-5}{\rm{eV}}$ and a decay constant $f_a=5.20\times10^{12}{\rm{GeV}}$ would emit a radio burst with a frequency of $\nu=6.05\rm{GHz}$ when its size is less than the critical radius $R_{cr}$. The burst has a duration of approximately a few milliseconds, as illustrated in Fig.~\ref{fig:number2}, and a peak luminosity of $1.60\times10^{42}\rm{erg/s}$. The observed frequency is $\nu(1+z)^{-1}$ due to cosmological redshift, which can range from 0.45 to 0.96 \cite{Thornton:2013iua}. The frequency and luminosity are roughly consistent with those of FRBs \cite{Petroff:2019tty}.
Refer to Fig.~\ref{fig:constraints} for the constraints on the parameter space of axions.
Therefore, the stimulated decay of collapsing dilute axion stars for ALPs with smaller values of the axion-photon coupling than QCD axions may account for some of the observed non-repeating FRBs.

\section{Conclusions}
The QCD axion or ALP is a prominent candidate for dark matter. Axions can collectively form a bound state known as an axion star through BEC. It is possible that a significant fraction of dark matter is composed of these axion stars. When axion stars surpass a critical mass due to accretion or merging, they undergo a collapse driven by self-interactions. For spherically symmetric clumps, the parametric resonance of photons is generally absent when the QCD axion-photon coupling is at conventional levels. Still, it can occur with axions possessing moderately higher couplings. When the star's size approaches the Compton wavelength of the axion, $2\pi/m_\phi$, the axions undergo annihilations, transitioning into relativistic states. Nonetheless, our results indicate that a collapsing QCD axion star with an axion mass $m_\phi\gtrsim10^{-6}{\rm{eV}}$ will trigger stimulated decay without the need for an axion-photon coupling exceeding the original KSVZ value by two orders of magnitude, rather than generating relativistic axions. This radio burst, lasting for about 0.1 seconds, is potentially detectable by radio telescopes like FAST or SKA. In addition, the collapsing axion star for ALPs with $m_\phi=5\times10^{-5}{\rm{eV}}$ and decay constant $f_a=5.20\times10^{12}{\rm{GeV}}$ would emit a radio burst with a frequency of $\nu=6.05\rm{GHz}$ when its size is less than the critical radius $R_{cr}\sim24\pi\Gamma_\phi N_\phi/m_\phi^2$, with a duration of approximately a few milliseconds and a peak luminosity of $1.60\times10^{42}\rm{erg/s}$. The frequency and luminosity are roughly consistent with the characteristics of FRBs. This could indicate the existence of axion stars and may have implications for laboratory experiments aimed at detecting axions, shedding light on the nature of dark matter.

\section{Acknowledgments}

This work was supported by National Natural Science Foundation of China under Grant No. 11947031 and East China University of Technology Research Foundation for Advanced Talents under Grant No. DHBK2019206.

\end{document}